# Driving magnetic order in a manganite by ultrafast lattice excitation

M. Först[1*], R.I. Tobey[2*], S. Wall[3*], H. Bromberger[1], V. Khanna[1,4,11], A.L. Cavalieri[1], Y.-D. Chuang[5], W.S. Lee[6], R. Moore[6], W.F. Schlotter[7], J.J. Turner[7], O. Krupin[8], M. Trigo[9], J.C. Mitchell[10], S.S. Dhesi[11], J.P. Hill[2], A. Cavalleri[1,4#]

[1]Max-Planck Department for Structural Dynamics, Center for Free Electron Laser Science, University of Hamburg, Germany

[2]Condensed Matter Physics and Materials Science Department, Brookhaven National Laboratory, Upton, NY

[3]Fritz-Haber-Institute of the Max Planck Society, Berlin, Germany

[4]Department of Physics, Clarendon Laboratory, University of Oxford, United Kingdom

[5]Advanced Light Source, Lawrence Berkeley Laboratory, Berkeley, CA

[6] SIMES, SLAC National Accelerator Laboratory and Stanford University, Menlo Park, CA

[7] Linac Coherent Light Source, SLAC National Accelerator Laboratory, Menlo Park, CA

[8]European XFEL, Hamburg, Germany

[9]PULSE, SLAC National Accelerator Laboratory, Menlo Park, CA

[10] Materials Science Division, Argonne National Laboratory, Argonne, IL

[11] Diamond Light Source, Chilton, Didcot, Oxfordshire, United Kingdom.

**Optical control of magnetism, of interest for high-speed data processing and storage, has only been demonstrated with near-infrared excitation to date [1,2,3]. However, in absorbing materials, such high photon energies can lead to significant dissipation, making switch back times long and miniaturization challenging. In manganites, magnetism is directly coupled to the lattice, as evidenced by the response to external and chemical pressure [4,5], or to ferroelectric polarization [6,7,8]. Here, femtosecond mid-infrared pulses are used to excite the lattice [9,10] in $La_{0.5}Sr_{1.5}MnO_4$ and the dynamics of electronic order are measured by femtosecond resonant soft x-ray scattering with an x-ray free electron laser. We observe that magnetic and orbital orders are reduced by excitation of the lattice. This process, which occurs within few**

**picoseconds, is interpreted as relaxation of the complex charge-orbital-spin structure following a *displacive exchange quench* – a prompt shift in the equilibrium value of the magnetic and orbital order parameters after the lattice has been distorted. A microscopic picture of the underlying unidirectional lattice displacement is proposed, based on nonlinear rectification of the directly-excited vibrational field, as analyzed in the specific lattice symmetry of $La_{0.5}Sr_{1.5}MnO_4$. Control of magnetism through ultrafast lattice excitation has important analogies to the multiferroic effect and may serve as a new paradigm for high-speed optomagnetism.**

At low-temperatures ($T < T_N$=110 K $< T_{CO/OO}$=220 K), single-layer $La_{0.5}Sr_{1.5}MnO_4$ exhibits CE-type charge, spin and orbital order, characterized by in-plane "zig-zag" ferromagnetic chains. These chains are antiferromagnetically coupled with one another, in and out of plane [11,12,13]. Resonant soft X-ray diffraction is directly sensitive to this spin and orbital order, when the incident photon energy is tuned to the $2p \rightarrow 3d$ transitions (Mn $L_{2,3}$ edges), and provides both momentum-dependent and spectroscopic information [14,15]. Figure 1(a) shows the top-view of the relevant scattering geometries at the orbital (¼ ¼ 0) and the magnetic (¼ ¼ ½) wave vectors for a $La_{0.5}Sr_{1.5}MnO_4$ crystal cut with a (110) surface normal. Static energy scans at the Mn $L_3$ and $L_2$ edges, also displayed, are in agreement with the literature [16, 17].

In our experiments, $La_{0.5}Sr_{1.5}MnO_4$ was held at a base temperature of 25 K, and excited by 130-fs 1.2-mJ/cm$^2$ mid-infrared pulses, obtained by difference-frequency mixing of the signal and idler output from an infrared optical parametric amplifier. These pulses, polarized in the *ab*-plane of the sample, were tuned at a centre wavelength of 13.5 µm (92 meV) with a 4.5-µm FWHM bandwidth that covered the 16-µm (78 meV) Mn-O stretching vibration [18]. Mid infrared pulses at this wavelength, even if slightly detuned to the red of the $La_{0.5}Sr_{1.5}MnO_4$ phonon resonance, were previously shown to effectively excite the lattice and perturb electronic order [19].

The excitation pulse train was synchronized to the 60-Hz repetition rate of the LCLS, an X-ray free electron laser (FEL) [20], which was operated at 640 eV in the 40-pC low-charge mode with sub-30 fs pulse duration. Further filtering by a soft x-ray monochromator allowed the photon energy to be set to the Mn $L_3$ resonance. Femtosecond Resonant Soft X-ray Diffraction [21,22] was then used to probe orbital and magnetic order dynamics with 250-fs time resolution, limited by the timing jitter between the mid-infrared and x-ray pulses. The diffracted light was detected for each time delay with a fast-readout CCD camera, recording all images individually at 60 Hz. The diffraction spots for the two different scattering geometries are shown in Fig. 1(b) and (c) after averaging over 70 images. These images are oriented in such a way that the scattering plane of the experiment is horizontal. For an analysis of transient changes in spot intensity, position, and width, CCD images at each time delay were projected in both, horizontal and vertical directions and fitted with Lorentzian profiles on top of a linear background (see also Fig. 1).

The temporal evolution of the integrated diffraction spot intensity at the magnetic (¼ ¼ ½) wave vector, obtained from the fits as described above, is reported in Figure 2(a). Diffraction was reduced by 8%, with a single time constant of 12.2 ps. For comparison, we display the significantly faster response response measured after excitation with 5-mJ/cm$^2$ pulses at 800-nm wavelength [23,24,25], which reveals a prompt collapse of magnetic order on the 250 fs time resolution of the experiment. This observation of different timescales is evidence that lattice driven magnetic disordering must follow a different physical path than for electronic excitation in the near infrared.

In Figure 2(b), we display the orbital order dynamics measured at the (¼ ¼ 0) diffraction peak after mid-infrared excitation. The transient responses of the orbital and magnetic peaks differ both in the timescale and amplitude, with the orbital order only reduced by only 3% with a single-exponential decay time of 6.3 ps. We note that this lattice-driven orbital disordering is slower than was observed previously by time-dependent optical birefringence [19]. However, time dependent optical birefringence, proportional to the orbital order parameter squared in equilibrium [18], is a less direct method than the resonant x-ray diffraction used here.

Throughout these dynamics, we see no transient change in the position and width of the scattered diffraction spots for either order and conclude that the correlation lengths are not perturbed. This is shown in Figure 2(c) where we exemplarily plot the transient width of the magnetic diffraction spot together with its peak position. The latter is constant within $< 1 \times 10^{-5}$ (calculated standard deviation). For a thermal expansion coefficient of $6 \times 10^{-6}\,\text{K}^{-1}$ [26] this corresponds to a temperature increase less than 2 K, indicative of the low dissipation associated with this vibrational excitation.

Recently [27], lattice dynamics observed in $La_{0.7}Sr_{0.3}MnO_3$ following mid-infrared excitation have been explained in terms of *Ionic Raman scattering* (IRS) [28,29]. The key step of IRS is the direct resonant excitation of a large amplitude infrared-active vibration, whose field is rectified through the second-order lattice polarizability – a process analogous to rectification in nonlinear optics. If $Q_{IR}$ is the amplitude of the normal coordinate of the infrared-active vibration, its rectification leads to a half cycle force field along the coordinate defined by the product symmetry group $Q_{IR} \cdot Q_{IR}$. Importantly, this half cycle field tends a finite area, resulting in a displacive force that leaves the lattice in a new position at the end of the pulse. The displacement has amplitude proportional to $Q^2_{IR}$ and Raman symmetry [27].

To understand the dynamic crystallographic problem in $La_{0.5}Sr_{1.5}MnO_4$, we show a schematic picture of its low-temperature charge- and orbital-order unit cell in Figure 3(a) [30]. This structure belongs to the $D_{2h}$ point group. The 630-cm$^{-1}$ (16-µm) stretching mode excited by the mid-infrared light field [18,31] has $B_{2u}$ symmetry (see Figure 3(b)) [32]. The symmetry of the rectified field belongs to the product group $B_{2u} \otimes B_{2u} = A_g$, among which we find the Raman-active Jahn-Teller mode depicted in Fig. 3(c). Thus, according to the IRS model, rectification of the mid-infrared mode is able to relax the cooperative Jahn-Teller distortion, which has no infrared activity and thus cannot be driven directly by mid-infrared excitation. Importantly, the Jahn-Teller mode shown in Fig. 3(c) relaxes the splitting between crystal field levels and reduces the ordering of the orbitals. In turn, this weakens the exchange interaction that stabilizes the CE-type order and would thus lead to a smaller equilibrium magnetization, or to a lower equivalent Neel temperature.

We stress that in contrast to the case of $La_{0.7}Sr_{0.3}MnO_3$ [27], in which the envelope of the infrared-active $E_u$ mode drives a low-frequency 1.2-THz rotational ($E_g$) mode impulsively, the Jahn-Teller $A_g$ mode has a higher frequency (15 THz) than the inverse 130-fs envelope of the infrared-active mode. Thus, in $La_{0.5}Sr_{1.5}MnO_4$, the $A_g$ distortion is driven adiabatically within 130 fs toward its distorted position, without coherent oscillations.

In Figure 4, we present a caricature of the magnetic and orbital order melting process. Within the 130-fs timescale of the mid-infrared excitation, the equilibrium positions of both order parameters are rapidly quenched to a new value, i.e. the minimum of the free energy surface is displaced within the magnetic and orbital order parameter plane. Due to this shift, being prompt compared to the response time of either order parameter, the system starts to rearrange from a state defined by the unperturbed free energy surface. An effective force is then exerted onto the spin and orbital degrees of freedom, driving a complex relaxation toward the bottom of the new free energy surface along the asymmetric gradient sketched in the figure. We refer to this prompt shift in the free energy surface as a *displacive exchange quench*.

Following this picture, the measured timescales for the two order parameters to settle to their new values are significantly different from one another, with the orbital order approximately twice as fast for the same excitation fluence. Despite the complexity of the non-equilibrium relaxation process after the quench, a few considerations aid our understanding of this difference. The loss of orbital orientation involves only rearrangements in the $e_g$ electrons, thus only one charge (or less) on each $Mn^{3+}$ site, and no significant change in spin momentum *per se*. We expect this effect to occur more rapidly. On the other hand, the loss of antiferromagnetic spin order, for which one $e_g$ and three $t_{2g}$ spins rotate, requires a significant exchange of spin angular momentum, and thus has significant inertia.

The mechanism of this displacive exchange quench following nonlinear lattice excitation qualitatively explains the disordering of the system, although the microscopic pathway and the non-equilibrium physics at play are not clear. Also, the recovery of the disordered state to thermal equilibrium on longer time requires further clarification. An important effect to be understood is related to the coupling to other degrees of freedom after the exchange quench, including the fate of the angular momentum and the entropy increase that follows coupling to the thermodynamic bath. An important goal with great potential for applications would be to find routes to reverse the sign of the Jahn-Teller distortion with a second pulse after the magnetic and the orbital orders have relaxed into their new equilibrium values. This would only be possible if the coupling to the bath was small and the generation of entropy minimal.

In summary, we have shown that direct excitation of the lattice in a manganite using high intensity mid-infrared pulses melts the spin order, an effect that can only be measured directly using a direct femtosecond x-ray probe. We explain this process by considering the ionic Raman scattering mechanism and the excitation of the $A_g$ Jahn-Teller motion through lattice nonlinearities, which is posited to lead to a displacive force on the spin and orbital order. Control of magnetism through direct distortion of the lattice may, in appropriate excitation geometries, lend itself to bi-directional switching, of interest for high-speed data processing applications.

## ACKNOWLEDGEMENT

The authors acknowledge stimulating conversations with S.B. Wilkins and R. Merlin.

This work was funded by the Max Planck Society through institutional support for the Max Planck Research Group for Structural Dynamics at the University of Hamburg. Portions of this research were carried out on the SXR Instrument at the Linac Coherent Light Source (LCLS), a division of SLAC National Accelerator Laboratory and an Office of Science user facility operated by Stanford University for the U.S. Department of Energy. The SXR Instrument is funded by a consortium whose membership includes the LCLS, Stanford University through the Stanford Institute for Materials Energy Sciences (SIMES), Lawrence Berkeley National Laboratory (LBNL), University of Hamburg through the BMBF priority program FSP 301, and the Center for Free Electron Laser Science (CFEL).

Work performed at Brookhaven was supported by US Department of Energy, Division of Materials Science under contract no. DE-AC02-98CH10886. Work at Argonne is supported under Contract No. DE-AC02-06CH11357 by UChicago Argonne, LLC, Operator of Argonne National Laboratory, a U.S. Department of Energy Office of Science Laboratory.

S. Wall acknowledges support from the Alexander von Humboldt Foundation.

**FIGURE CAPTIONS**

**Fig.1: Femtosecond resonant soft X-ray diffraction.** (a) Top-view of the pump-probe scheme and scattering geometries for the orbital (¼ ¼ 0) and the magnetic (¼ ¼ ½) diffraction peaks, measured on a (110) surface normal crystal of $La_{0.5}Sr_{1.5}MnO_4$. The Bragg planes are indicated as parallel lines; polarizations of the LCLS X-ray beam and the mid-infrared light are horizontal and vertical, respectively. Also shown are the energy dependencies measured at beamline X1A2, National Synchroton Light Source with a 1.5 eV resolution, comparable to the energy resolution at the free electron laser. Panels (b) and (c) show the diffracted spots, recorded over 180×180 pixels of a fast CCD at LCLS. Images represent the average of 70 shots of the FEL. For the analysis of transient changes in the scattering, these images were projected in both, horizontal and vertical, directions and fitted with Lorentzian profiles (red solid lines) on top of a linear background.

**Fig.2: Lattice induced magnetization dynamics.** (a) Time-resolved changes of the intensity of the scattered X-ray spot at the magnetic (¼ ¼ ½) diffraction peak for vibrational excitation at 13.8 µm (mid-infrared) and photo-doping at 800 nm (near-infrared) pump wavelengths. Excitation fluences are 1.2 and 5 mJ/cm$^2$, respectively. Data were obtained as described in the text. (b) Comparison of the melting of the orbital and magnetic order for 1.2-mJ/cm$^2$ mid-infrared excitation. The red lines are single exponential fits to the data, with time constants of 6.3 ps and 12.2 ps, respectively. (c) Transient change of the scattering angle 2θ and the width *w* of the diffracted x-ray spot measured for the magnetic wave vector following the mid-infrared excitation. These data were obtained by projecting the diffraction spot intensities in the vertical direction (see also Fig. 1(b)) and fitting the resulting 1D-profiles in the horizontal direction (scattering plane).

**Fig.3: Ionic Raman scattering.** (a) Schematic of charge and orbital order pattern in the *ab*-plane of $La_{0.5}Sr_{1.5}MnO_4$, together with the orbital order unit cell (thick light-blue line). Displacements of Mn

and O atoms associated with (b) the resonantly driven IR-active $B_{2u}$ mode and (c) the Raman-active $A_g$ mode driven through *Ionic Raman scattering* are also shown. The latter mode relaxes the Jahn-Teller distortions to affect the exchange interaction, and thus the electronic and magnetic state of the system.

**Fig.4: Displacive exchange quench model.** Excitation of the lattice reduces the Jahn-Teller distortion as in Figure 3(b), resulting in an effective shift in the equilibrium value of the orbital and magnetic order parameters. The effective force on each order parameter is given by the projection of the gradient along each axis.

**FIGURES**

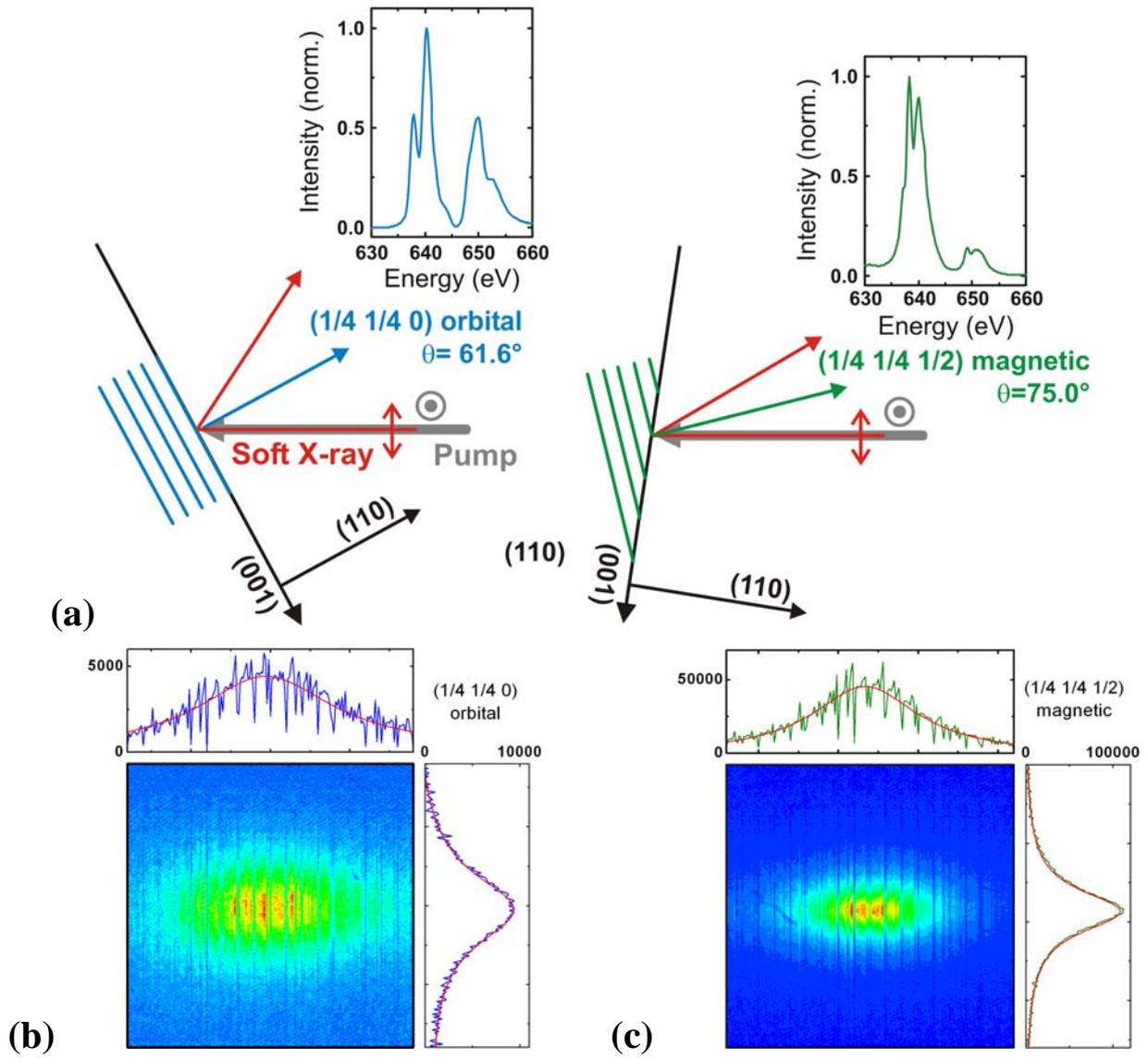

Figure 1

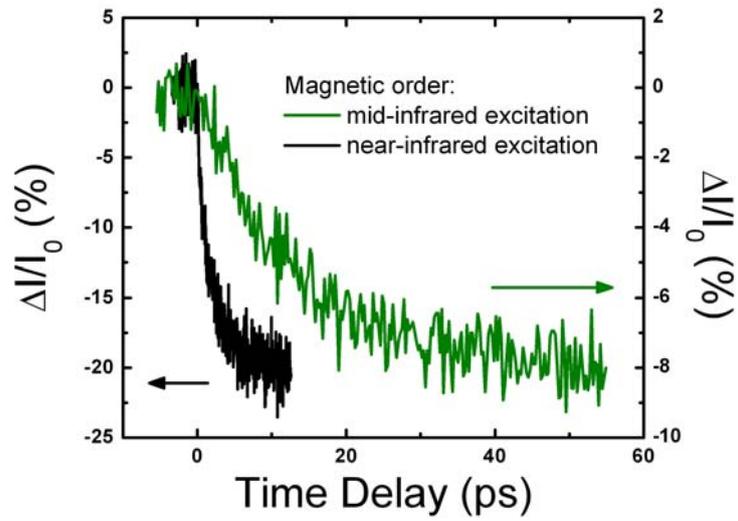

**Figure 2a**

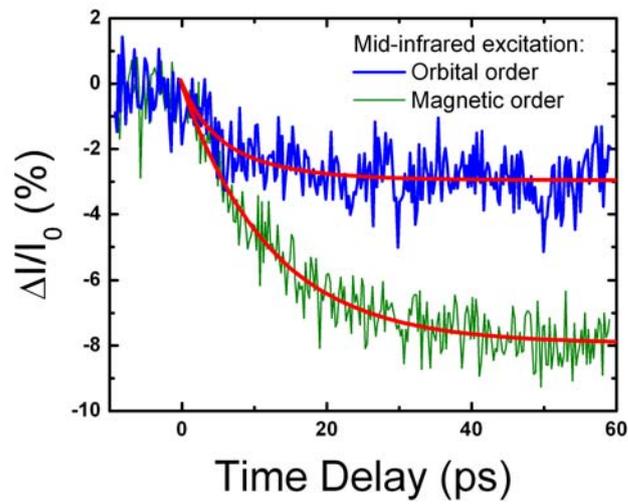

**Figure 2b**

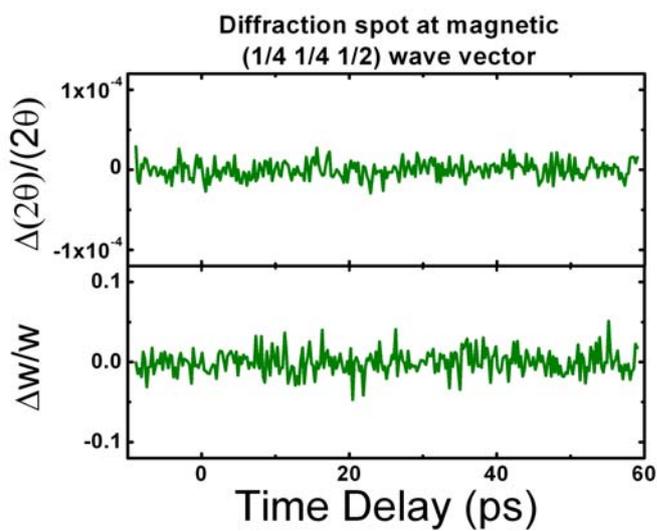

**Figure 2c**

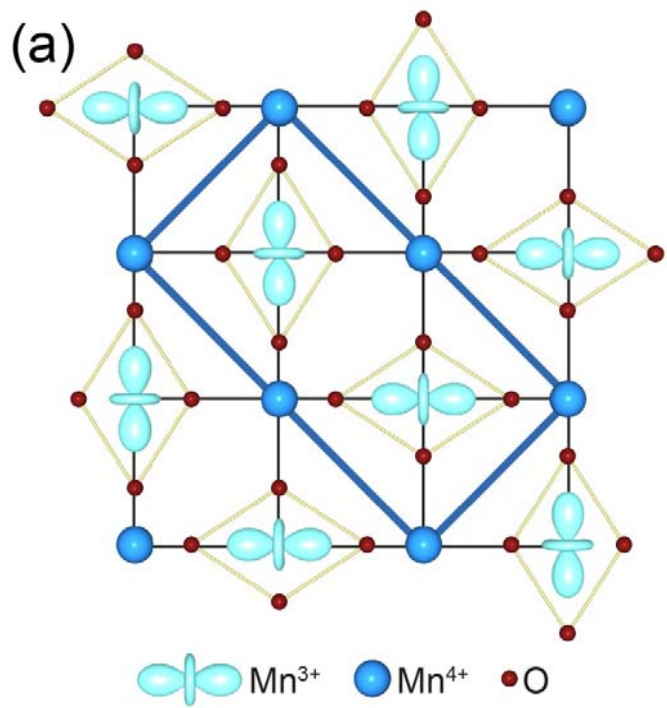

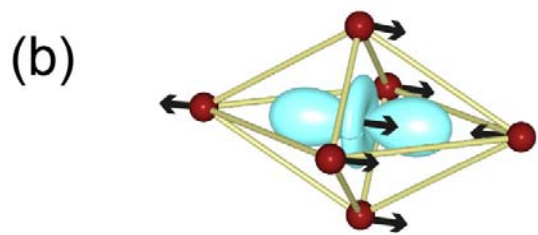

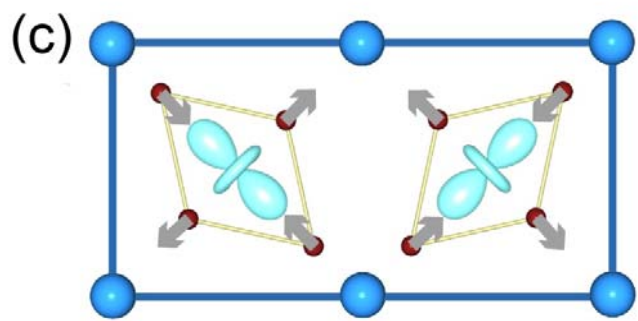

**Figure 3**

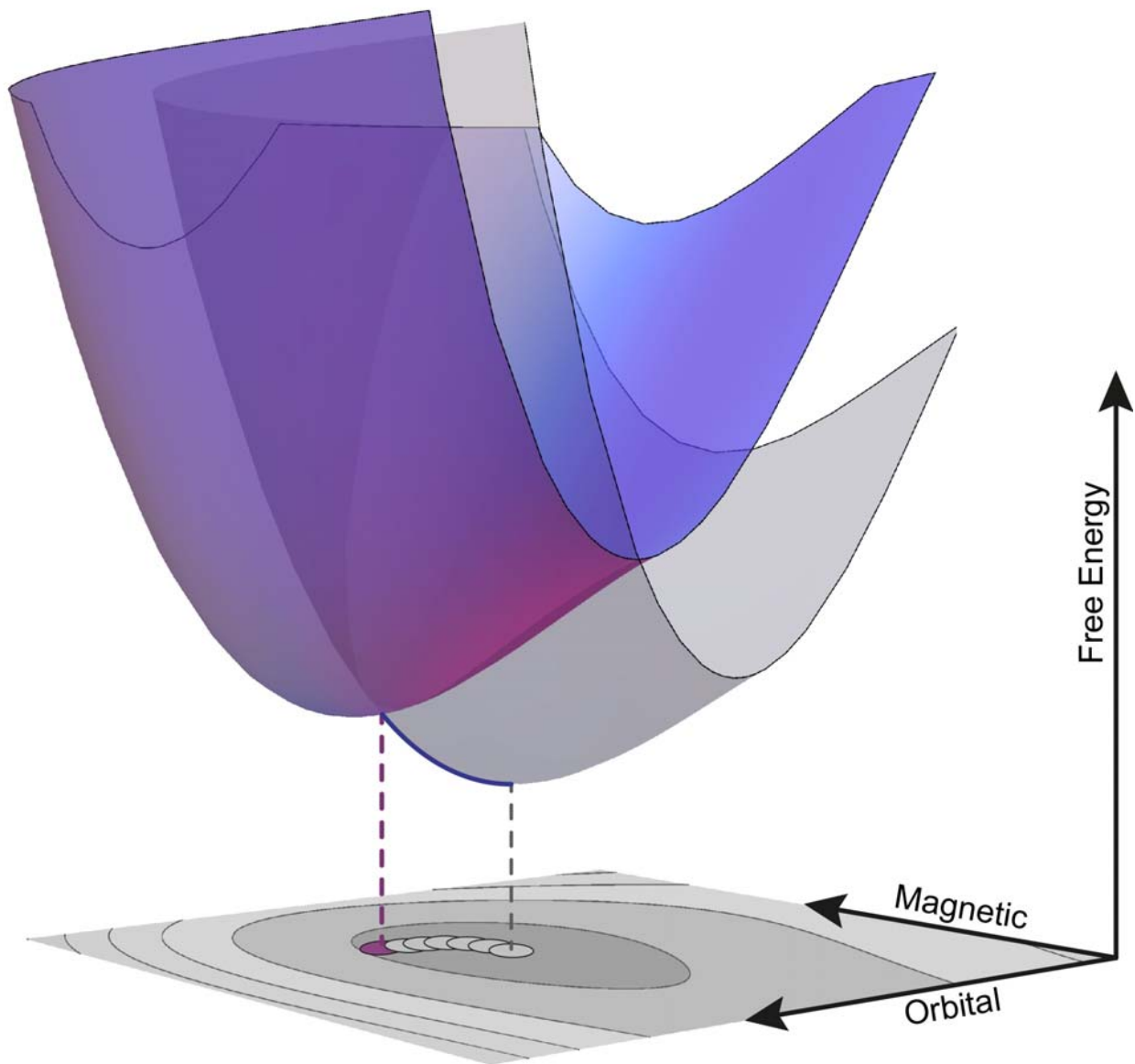

**Figure 4**